# Rapid voltage sensing with single nanorods via the quantum confined Stark effect

Omri Bar-Elli[*], Dan Steinitz[*], Gaoling Yang[*], Ron Tenne[*], Anastasia Ludwig[§], Yung Kuo[+], Antoine Triller[§], Shimon Weiss[+‡], Dan Oron[*]

[*]Department of Physics of Complex Systems, Weizmann Institute of Science, Rehovot 76100, Israel

[§]LEcole Normale Superieure, Institute of Biologie (IBENS), Paris Sciences et Lettres (PSL), CNRS UMR 8197, Inserm 1024, 46 rue d'Ulm, Paris 75005, France

[+]Department of Chemistry and Biochemistry, Department of Physiology, and California NanoSystems Institute, University of California Los Angeles, Los Angeles, California

[‡]Department of Physics, Institute for Nanotechnology and Advanced Materials, Bar-Ilan University, Ramat-Gan, 52900, Israel

**Abstract**

Properly designed colloidal semiconductor quantum dots (QDs) have already been shown to exhibit high sensitivity to external electric fields via the quantum confined Stark effect (QCSE). Yet, detection of the characteristic spectral shifts associated with the effect of QCSE has traditionally been painstakingly slow, dramatically limiting the sensitivity of these QD sensors to fast transients. We experimentally demonstrate a new detection scheme designed at achieving shot-noise limited sensitivity to emission wavelength shifts in QDs, showing feasibility for their use as local electric field sensors on the millisecond time scale. This regime of operation is already potentially suitable for detection of single action potentials in neurons at a high spatial resolution.

Fluorescent markers sensitive to the electric field in their local environment have found extensive use in biological studies.[1–3] This is particularly true for spatially mapping activity in neural networks, where classical electrophysiological approaches do not allow probing of the entire neural circuit. Most observations of neuronal activity are based on calcium imaging,[4] an indirect proxy of membrane potential dynamics due to the slow kinetics of calcium transients. Lately, considerable efforts have been invested in development of voltage-sensitive organic dyes[5] (VSDs) and genetically encoded voltage-sensitive proteins.[6] Although these tools are likely to become essential for studying the brain, their performance is still lacking either the temporal or the spatial precision needed for simultaneous optical recording of action potentials (APs) from a large number of neurons inside the brain of a live, behaving animal. The current best performing VSDs are based on photoinduced electron transfer between an electron rich quencher and an organic fluorophore.[2,5,7–9] This leads to a modulation of the fluorescence intensity as a function of the electric field. Yet, these markers still suffer from poor membrane retention, toxicity, membrane capacitance perturbation, photobleaching, and are incapable of resolving small features in neuronal membranes.[2,5,9]

Highly fluorescent semiconductor nanoparticles, known as quantum dots (QDs), were recently suggested as an alternative to the classical VSDs[10–14] for direct detection of electric fields via the use of the quantum confined Stark effect (QCSE). First observed in quantum wells,[15] the QCSE leads to a shift in the luminescence center wavelength of the QD.[15,16] QCSE in QDs was first observed over two decades ago,[17,18] but was usually characterized in the ensemble.[19] Additionally, QCSE from single QDs is easier to observe at low temperatures, where thermal broadening of the emission linewidth of QDs is negligible. Observation of QCSE induced spectral shifts at room temperature was thought to be challenging due to the presence of stochastic meandering of the emission center wavelength known as spectral diffusion.[17] Yet, emission wavelength shifts of several nanometers were recently reported at room temperature on single particles.[16] The characteristic electric fields due to APs, of the order of hundreds of KV/cm represent the perturbative limit of QCSE described above. Thus, optical phenomena observed under much stronger fields are irrelevant.[20]

QD markers exhibit several qualities that make them potentially favorable over their organic counter parts. Owing to their nanoscale volume, it is possible to embed QDs-based sensors at different locations along cell membranes,[14] opening the possibility for sensing and resolving the local electric field in a sub-diffraction volume by using a single reporter QD. These sensors could also potentially be targeted to synapses by conjugation with specific peptides recognizing postsynaptic scaffolds, such as PSD95 and gephyrin.[21–23] Furthermore, fluorescence quantum yield and photo-stability of QDs are significantly higher,[24] qualities that can be preserved even in biological media,[14] allowing for high photon flux from a single emitter over prolonged observation times. This is crucial for biological applications taking into account toxicity due to intense illumination for live samples.[25] In striking difference with standard VSDs and protein-based sensors, QD-based sensors can be used at very low concentrations without altering electrical properties of the membrane.[26,27] These unique advantages of QD-based sensors open up a new avenue for the super-resolution voltage imaging in living cells. However, to be useful in realistic neuroscience applications, QCSE has to be observed on the time scale of an AP, corresponding to a sub-millisecond time scale,[28] in stark contrast to the characteristic one second time resolution of typical studies of either QCSE or spectral diffusion.[29] The main goal of the following work is to examine whether detection of single particle spectral shifts on a millisecond time scale is possible, and whether these shifts can be measured down to the limit afforded by the shot noise, the noise limit in any classical physical system due to the probabilistic nature of the measurement process.

The response of QDs to an external electric field is usually described using perturbation theory.[15] For a symmetric quantum well the first order correction to the emission wavelength vanishes and thus the spectral shift has a second order dependence on the external field. As a result, in a spatially symmetric system an external electric field leads to a decrease in the band gap energy and a concomitant red shift of the emission spectrum, quadratic with the field amplitude. For an asymmetric well, in which the exciton has a permanent electric dipole, the energy shift is linear in the external field resulting in either a blueshift or a redshift of the spectrum, depending on the relative orientation of the exciton dipole moment and the external field. Indeed, both blueshifts and redshifts were predicted[11,16] and observed for asymmetric type-II seeded nanorods.[16] Neural membrane resting potential limits the application of symmetric wells for neuro imaging. While the potential difference between the resting potential (~-70mV) and the AP spike (~+30mV) is large,[28] a symmetric structure would exhibit only a small redshift due to the difference of the absolute values (~40mV). In contrast, an asymmetric structure, responding in a linear fashion to voltage change can exhibit a large spectral shift relating to the sum of absolute values (~100mV). In addition, since the electric field also modifies the spatial wavefunction of the electron and hole and influences their overlap one expects the radiative lifetime to be affected. In general, an increase in the emission energy should be accompanied by an increased spatial hole and electron wavefunction overlap and a shortening of the radiative lifetime.[11]

Several physical quantities can, in principle, be utilized for detecting changes in the local electric field. Most voltage sensors exhibit a change in the emission intensity due to the presence of the field. This is easily detected over a wide illumination area and is thus compatible with standard neuronal imaging setups. While this effect was measured in QDs[12,16] it is expected, to be rather small for high quantum efficiency QDs, where exciton recombination is dominated by radiative decay as is necessary for fast detection.[16] Perhaps the most intuitive signature that can be measured is the spectral shift, but lifetime imaging is also a possibility. Both of these approaches have been explored for VSDs[30–32] and have generally been found to be inferior to imaging using intensity changes.[2,33] We present both of these voltage dependent signals for QDs based sensors. A more detailed theoretical discussion of the various detection schemes can be found elsewhere.[10,11]

From the above considerations, it is clear that asymmetric QDs should be more sensitive to the effect of electric fields, as has recently been shown both theoretically[11] and experimentally.[16] Here, we have chosen to work with type-II seeded nanorods (NRs) consisting of a spherical ZnSe core overcoated by a CdS rod, as seen in the TEM image of Figure 1a and illustrated in Figure 1c.

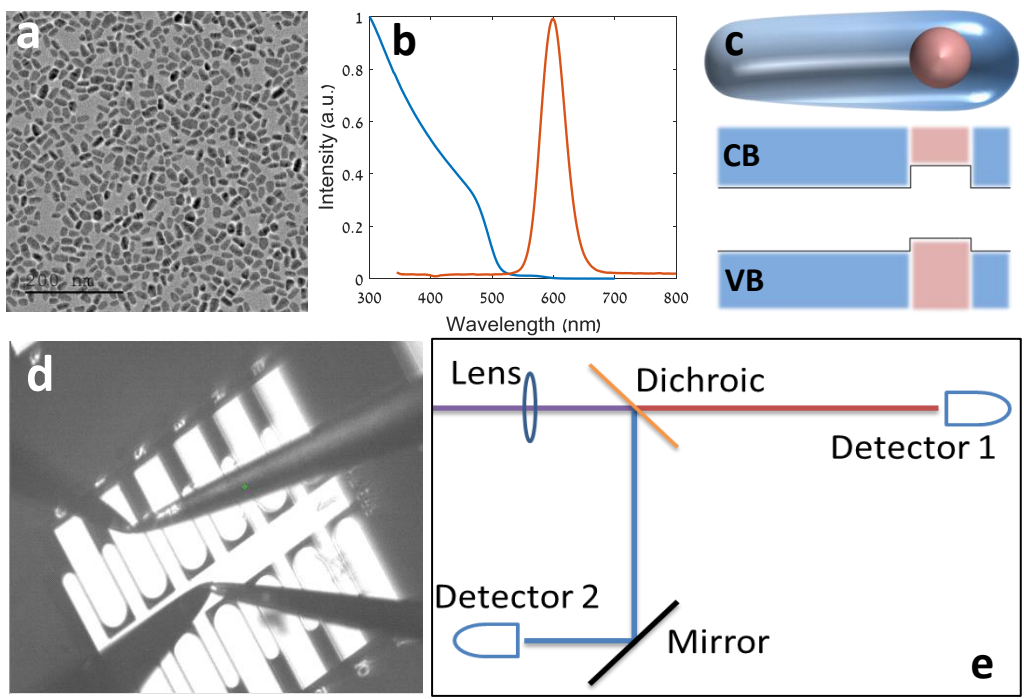

Figure 1 – (a) TEM image of the Type II ZnSe/CdS NRs used in all experiments, scale bar is 200nm. (b) Extinction (blue) and photoluminescence (red) spectra of the NRs. (c) A cartoon depicting the NRs shape. The rod-shaped shell is asymmetric and the core is closer to the thicker edge. Below is the energy levels diagram of such type-II NRs. CB – conduction band, VB – valance band. (d) An overhead view of a cover slip fashioned with gold electrodes that were used to apply the electric field to the NRs, the probe needles are touching the ground (central) and one of the "live" electrodes. (e) The fluorescence detection scheme used whereby, a dichroic mirror splits the PL peak into two channels.

In these particles, the core is not centered in the shell but rather close to the thicker edge of the rod.[34] Such particles have already been shown to exhibit sizable QCSE due to the separation of the charge carriers' wavefunctions, and a close to linear dependence on the field.[16] Synthesis of ZnSe/CdS nanorods was performed according to an adapted known procedure[34] yielding particles of 14.4±2.4nm in width and 24.7±3.7nm in length (Fig. S1), these have been shown to be an asymmetric type II structure.[35] The extinction and photoluminescence (PL) of the ensemble suspended in solution are given in Figure 1b.

To characterize the ability to rapidly detect electric field changes, we deposit NRs on a horizontal electrode array. Figure 1d shows a top view of a cover slip patterned with gold electrodes. The large central electrode has six protruding fingers on each side and the secondary electrodes are positioned between them. The gap between the electrodes is several microns, depending on the fabrication process. Voltage is applied using two metallic micromanipulator needles. The first, connected to the center electrode is grounded, whereas the second is connected to an amplified voltage source, which is modulated at 1KHz and with a duty cycle of 50%.

Single NRs are identified within the gaps between the electrodes by observing blinking in a camera image taken through the microscope objective. Emission from the NRs is collected through an oil-immersion objective and then split using a dichroic mirror (Fig. 1e), set to the center emission wavelength of the ensemble, onto two single photon detectors. The detection time stamps are measured and logged using a time-correlated single photon counting module (see methods).

The typical width of a single NR emission spectrum is on the order of tens of nanometers, thus detecting a shift of only a few nanometers may be challenging. The "balanced detector" depicted in Figure 1e is designed to provide maximal sensitivity to the emission spectral shifts. In addition, the use of high temporal resolution single photon detectors enables monitoring changes in the decay rate of the emission. Thus, three independent measurements are performed simultaneously: Intensity fluctuations known as ΔF, spectral shifts (Δλ), extracted from the intensity ratio of the two detectors, and lifetime variations (Δτ).

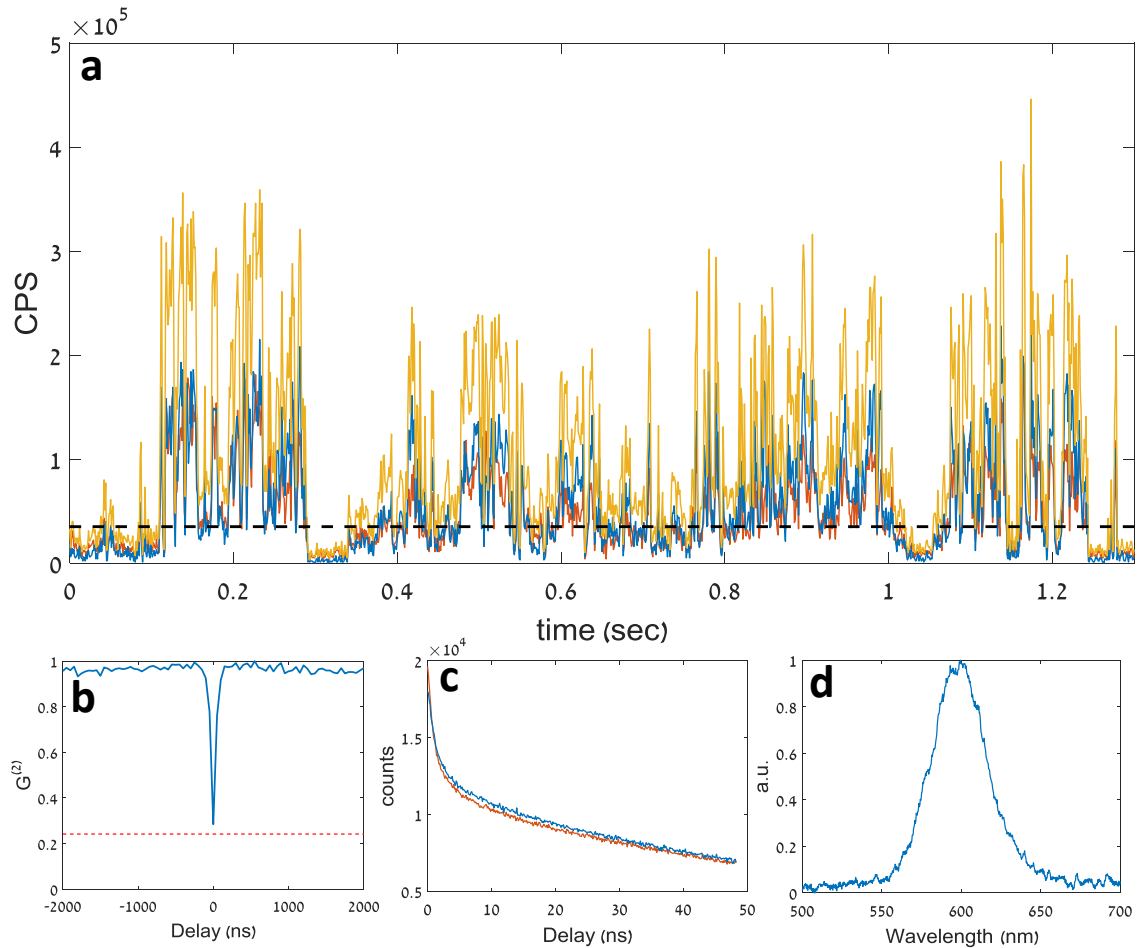

Figure 2 – (a) An example of a blinking trace (bin size - 1ms) of a single NR when detected by the setup depicted in Figure 1e. The intensity recorded in the reflection (blue) and transmission (red) channels are almost identical showing a 50:50 split of the emission peak by the dichroic. The sum of the signal from both channels (yellow) presents detection rates of up to 400kHz. The threshold chosen for this measurement is shown with a black dashed line. (b) Photon detection coincidences as a function of delay between the two channels. A dip is evident at zero delay where the correlation drops to 6% after correcting for the background indicating that the fluorescence is collected from a single emitter. An estimated background level is plotted in dashed red line. (c) Fluorescence decay in counts per second (CPS) as a function of time in the reflection (blue) and transmission (red) channels. (d) An example spectrum from a single NR (30s exposure).

A typical time trace of a single NR emission is presented in Figure 2a where the intensity recorded by each channel is approximately half of the total intensity. It is clear that the emitter fluctuates between a bright, a dark and, in some cases, a "grey" intermediate state.[36,37] This phenomenon, known as blinking, is further illustrated by examining the histogram of the trace (Fig. S2). Using the sum of the two channels a threshold is applied in order to eliminate dark "off" state periods from the analysis (Fig. S2).

QDs are, to a large degree, single photon emitters. After a single excitation cycle typically only a single photon is emitted even if multiple excitations occurred. This leads to an anti-correlation between detections in the two channels at times shorter than the radiative lifetime.[38,39] In Figure 2b such a correlation plot is given showing a significant anti-bunching dip at zero 6% after correcting for correlations due to background. Note, that for two uncorrelated emitters the anti-bunching at zero delay would be at least 50%, assuming similar intensities for both emitters. Using this method, we ensure further analysis is performed only on single NRs. Note, that estimating the anti-bunching is not essential for imaging but is rather used here to separate clustering effects.

A typical single NR spectrum is shown in Figure 2d; such measurements require long exposures (10s of seconds) making them unsuitable for detection of fast spectral shifts. Furthermore, to overcome the noise of the camera employed as the detector, spectra acquisition requires high excitation intensities that may bleach even the more stable NRs rendering them non-fluorescent.

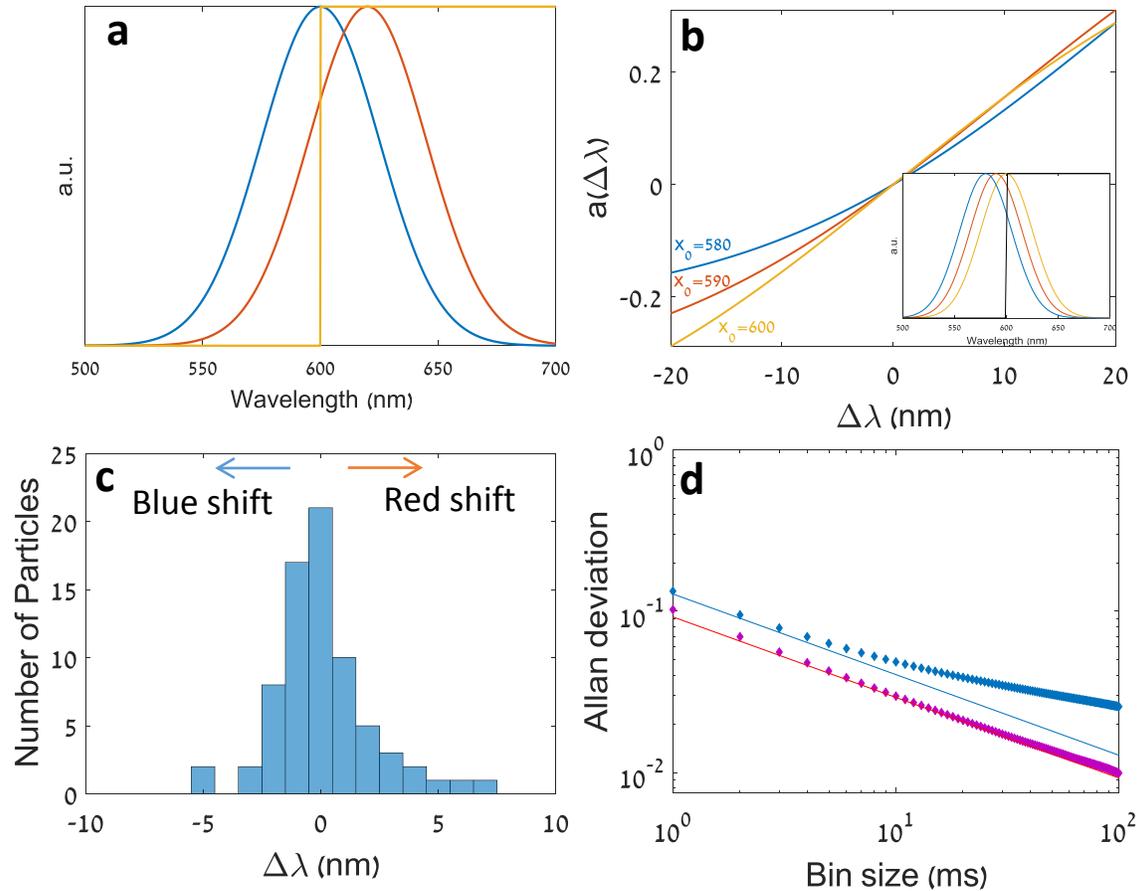

Figure 3 – (a) Simulated emission spectrum centered at $x_0$=600nm with σ=25nm (blue). The same spectrum shifted by Δλ=10nm while maintaining its width and area (red). A Step function at 600nm, simulating a dichroic mirror's transmission coefficient (yellow). (b) Three examples of the QCSE estimator ('a') plotted as a function of the wavelength shift (Δλ), where $x_0$ is the center of the original spectrum before voltage was applied; this is used to estimate Δλ from the data. The inset shows the three simulated spectra used and a step function (c) A histogram of 73 QCSE spectral shifts measured from 59 NRs under various electric fields equivalent to those in neuronal membranes. (d) Allan deviation of the Δλ estimator of two different NRs as a function of bin size (diamonds), compared with theoretical curves of shot noise (lines) when accounting for the count rate.

In order to extract the wavelength shift from the data collected in our setup, where the emission peak is split by wavelength into two channels, we used the following model, which is in line with previously discussed procedures.[11] The emission peak is modeled as a Gaussian function (Fig. 3a), whose width is determined from the single particle spectra that were collected and defined as σ=25nm for all calculations (see supplementary info). A step function is used to simulate a dichroic mirror with zero loss and 50% transmission at a chosen wavelength (typically 600-605nm). Next, a second, spectrally shifted, Gaussian is used to model a shift due to QCSE. One needs to define a quantity that reliably relates the actual spectral shift in nanometers and the measured intensities. While one can propose different estimators such as the intensity ratio between the two detection channels,[11] it is advantageous to use an estimator that has the same properties as the estimated quantity, in this case linearity with the electric field, we chose to use:

$$(1) \quad a(\Delta\lambda) = \frac{t_V}{t_V + r_V} - \frac{t_0}{t_0 + r_0} = \frac{t_V}{I_V} - \frac{t_0}{I_0}$$

Where *t* and *r* are the transmission and reflection intensities recorded in the two channels, respectively, and the subscript indicates for which Gaussian they were calculated, *0* for the original Gaussian and *V* for the shifted one. This quantity is sensitive to wavelength shifts but not to intensity fluctuations since each part is normalized to the total intensity recorded in the same time period. Three examples of this estimator are given in Fig 3b corresponding to three particles with a different center wavelength of emission. The sensitivity of the estimator and detection scheme in general, is given by the slope of this curve. One can observe that while this sensitivity is reduced for a particle whose spectrum is not centered on the dichroic mirror's transmission edge it remains relatively high even for shifts as large as 20nm. Clearly, maximal sensitivity is reached when the original spectrum is well centered on the dichroic or, alternatively, when the shift is in the direction of the dichroic cutoff wavelength.

Applying this model to experimental measurements is straightforward as it is possible to numerically extract the spectral shift from the value of the estimator (see supplementary info). It should be noted, however, that to determine whether or not an electric field was present does not require the aforementioned model. Rather, the model is used as a unit conversion method in order to present the results in units of wavelength. The results from 59 different NRs under various voltage modulation amplitudes are presented as a histogram in Figure 3c where it is evident that most NRs gave small, yet measurable, spectral shifts. This is most likely due to the shifts' dependence on the orientation[11,16] between the electric field and the NR. QCSE induced shifts are small when the electric field is perpendicular to the long dimension of the NR. The largest shifts measured are a red shift of +7.1nm and a blue shift of -5.5nm. Notably, the dependence of the spectral shift on the electric field amplitude is linear as expected for a type-II system, and a transition from redshift to blueshift is observed upon inversion of the field direction (Fig. S3).

To study the ability of this system to detect a transient applied voltage we used a simple voltage scheme, where a 1KHz, 50% duty cycle square wave was applied to the NRs (see methods). Each time bin of 1ms is divided into two halves, first when voltage is applied and second when no voltage is applied. The analysis is performed on each time bin separately such that the sensitivity and detection probability of a single short voltage pulse, similar in duration to an isolated AP is extracted. The estimator is calculated only for time bins that cross the defined intensity threshold. An example of the estimator distribution for a single particle is given in Figure S4. The mean and standard deviation of the distribution are used to report the spectral shift and its error.

To show that this measurement is indeed shot noise limited we calculate the error yielded by the model when only shot noise is considered (see supplementary info). Taking care to average only consecutive "on state" bins (see supplementary info), we present the Allan deviation of two different NRs in Figure 3d. A shot noise limited process would show a decrease in the error of $N^{-0.5}$, where $N$ is the number of photons or bins averaged over. Indeed, for short averaging time windows, the error in the estimator is only due to shot noise. At longer times, however, there is a decrease in the slope and the error does not improve as expected by shot noise. The averaging time for which the difference between the shot noise limit and the measured error varies among particle between few and tens of milliseconds, reaching more than 100ms for some particles. This type of deviation is expected in the presence of a slow "red" noise, such as the one induced here by spectral diffusion. However, as we show here, it has little effect on the ability to sense fast, millisecond scale, processes.

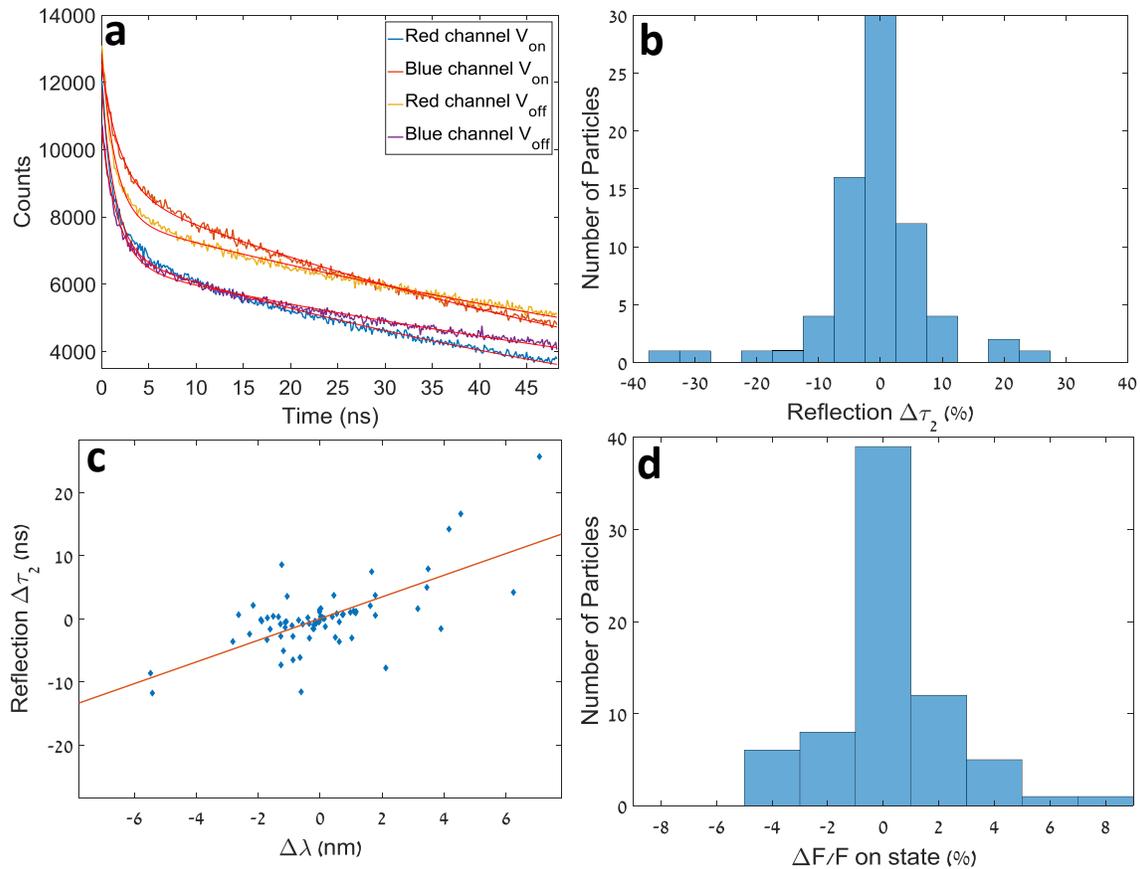

**Figure 4** – (a) An example of lifetime traces measured from a single particle in each channel while voltage was applied ($V_{on}$) and while it was not ($V_{off}$). A bi-exponential fit is plotted for each of the four curves yielding eight lifetime constants per particle. (b) A histogram of the longer lifetime variation ($\Delta\tau_2$) due to QCSE in the reflected channel. (c) A scatter plot depicting the correlation between spectral and lifetime shift. A trend line is shown as a guide for the eye. (d) A histogram of the intensity variations measured due to QCSE.

The use of avalanche photo diodes (APDs) affords high temporal resolution that enables an analysis of the fluorescence decay lifetime under an electric field at the single particle level. The collected data contains information of the lifetime of both parts of the spectrum (reflected blue part and transmitted red part). Each of these can be separated into periods in which voltage is applied ($V_{on}$) and those in which it is not ($V_{off}$). As may be seen in the example in Figure 4a, all four curves were well fitted with a bi-exponential fit yielding 8 lifetime constants ($\tau$). The lifetime variations are defined here as the difference in the decay lifetime within the same channel between $V_{on}$ and $V_{off}$ periods. Figure 4b shows all the measured lifetime changes of the long component due to QCSE in the reflected detection channel. Similar histograms are provided for the short component as well as for the transmission channel in Figure S5. As mentioned above, the spectral shift is to be accompanied by a change of the overlap between the wavefunctions of the hole and the electron changing accordingly the radiative lifetime decay. A positive correlation between the lifetime variation and the spectral shift is thus expected.[11] In Figure 4c this correlation is indeed evident: a blue shift (negative Δλ) is accompanied by a shortening of the lifetime (negative Δτ). Similar plots for both lifetime components channel are given in Figure S6. While we measure a clear lifetime change by using data acquired over tens of seconds, performing an analysis of the lifetime variation on a millisecond time scale, similar to the one we present for the Δλ analysis, is difficult. Normally, determining the lifetime from a small number of photons is done by taking the mean of their time of arrival. This approach is limited when the excitation repetition period is comparable to the radiative lifetime, which is the case here. Lowering the repetition rate of excitation would decrease the photon flux but should enable detection of transient electric fields using the lifetime data. As an alternative, NRs with shorter lifetimes may be used.

Determining the intensity fluctuation due to QCSE is done by examining the histogram of the blinking trace for periods of $V_{on}$ and $V_{off}$ separately (Fig. S2). The intensity change shows a weak correlation with the spectral shift (Fig. S2 inset) as expected for high quantum yield QDs.

While the correlations between Δλ, Δτ, ΔF are in agreement with the expected effect of QCSE, it is not possible to determine one by measuring the other as the degree of correlation is modest. Our data shows, for example, that the particle exhibiting the largest Δλ (7.1nm) also exhibits a negligible ΔF (0.2%). For the NR that exhibited the largest Δλ, we calculate a confidence level of 65% for detecting a 1.5ms square voltage pulse with a false positive probability smaller than 4% or, alternatively, a 50% confidence level with less than 1% false positive probability (Fig. S7).

We experimentally demonstrate a detection scheme for spectral shifts due to QCSE. We show that this scheme is sensitive to spectral shifts an order of magnitude smaller than the peak width. Moreover, we show our measurements are limited only by shot noise at short time scales. The detection scheme used enables simultaneous measurement of three effects applied voltage has on QDs, Δλ, Δτ, ΔF allowing for thorough evaluation of the correlations between them.

We conclude that the use of a 'balanced' detection scheme to observe spectral shifts yields superb sensitivity to voltage transients, enabling detection of a transient equivalent to a single AP with high levels of confidence especially when considering a realistic case of several NRs embedded in a single neuron membrane. Improving the detection capabilities further is possible with higher emission quantum efficiencies than the ones reported here. We find that measurement of spectral shifts yields better results even when compared to NRs with the strongest emission intensity modulation due to QCSE (where the underlying microscopic process is likely stochastic charging). Overall, our results point at the feasibility of using NR voltage sensors for rapid, wide-field voltage sensing with a high spatial resolution. Moreover, the 'balanced' detection scheme can be easily retrofitted to practically any commercial microscope using a standard low noise camera and a commercial imaging dichroic splitter.

**Methods**

A 470nm or a 510nm pulsed laser diode with 20MHz repetition rate (Edinburgh Instruments, EPL-470, EPL-510) were used for single NR excitation. The excitation laser was coupled into a microscope (Zeiss, Axiovert 200 inverted microscope) and focused using a high NA oil immersion objective (Zeiss, Plan Apochromat X63 NA 1.4). The epi-detected signal was filtered, using a dichroic mirror (Semrock, Di02-R488-25x36) and a long pass filter (Semrock, 488LP edge basic), and directed to a home built spectrally tunable balanced detection setup. A dichroic mirror (Semrock, Di02-R594) was used to split the emission peak from the NRs. Fine tuning of the dichroic cutoff wavelength to 600nm was enabled by changing the angle of incidence. Each part of the spectrally split signal was coupled into a multimode fiber and detected by an avalanche photodiode (Perkin Elmer SPCM-AQ4C) which was connected to a time-correlated single-photon counting (TCSPC) system (Picoquant, HydraHarp 400).

Single particle spectra were measured using a fiber coupled spectrometer (Princeton Instruments, Acton SP2300i) and a CCD camera (Princeton Instruments, Pixis256). A standard CCD camera (Thorlabs, DCC1645C) was used in top view imaging for positioning of the voltage probes.

Cover slips (#1.0, 25mm diameter) were prepared with gold microelectrodes (Fig. 1d, see supplementary info for further details on the preparation of electrodes) using standard clean room photolithography procedures. NR samples were diluted in 4% PMMA (Sigma Aldrich) in toluene and spin coated at 3000rpm. Probe positioners (Cascade Microtech, DPP-105-M-Al-S) were used to apply voltage to a chosen electrode pair, a multi-meter was used to measure the resistance between the electrodes to ensure there was indeed no short. Voltage was supplied from an amplifier (TREK, 2205) fed by a delay generator (Stanford Research Systems, DG645). A synchronized trigger was directed from the delay generator to the TCSPC to mark the beginning of each voltage cycle. In all experiments, the voltage applied was a 1kHz 50% duty cycle square wave of amplitude 50-100V after amplification. Producing voltage pulses of 0.5ms. Considering an ideal plate capacitor approximation, the electric field would be 125-400kV/cm depending on the inter electrode gap. In practice, fields are expected to be lower as this is far from an ideal plate capacitor. This field amplitude is comparable to the fields that exist in neuronal membranes. The voltage was applied to the central electrode (typically the ground) and an adjacent 'finger' electrode. To reverse the voltage in the same measurement the connections to the probes were switched.

**Acknowledgements**

The authors would like to thank K. Park for helpful discussions. The authors gratefully acknowledge funding by the Human Frontiers Science Project RGP0061/2015 and the European Research Council advanced grant NVS 669941. AL acknowledges support from the Marie Curie Individual Fellowship NanoVoltSens 752019. DO acknowledges support from the Crown center of photonics and the Israeli centers of research excellence program. SW acknowledges support by DARPA Fund #D14PC00141 and by the U.S. Department of Energy Office of Science, Office of Biological and Environmental Research program under Award Number DE-FC02-02ER63421

# Supplementary information file for

## Shot noise limited voltage sensing with single nanorods via the quantum confined Stark effect

Omri Bar-Elli[*], Dan Steinitz[*], Gaoling Yang[*], Ron Tenne[*], Anastasia Ludwig[§], Yung Kuo[+], Antoine Triller[§], Shimon Weiss[+‡], Dan Oron[*]

**Details of the collection setup**

As explained in the methods section, a dichroic mirror is used to split the emission spectrum of a single nanorod (NR) into two independent detectors. Figure 3b of the main text illustrates the importance of splitting the spectrum in half or as close as possible. Due to size variations between NRs their emission peak is shifted with respect to one another. To overcome this issue, we designed a tunable wavelength beam splitter. Dichroic mirrors are dielectric filters composed of thin layers of materials with different refractive indices. Each surface introduces a reflected wave and the interference of the reflection sums up into a wavelength dependent reflection coefficient. Owing to this design, the transmission (and reflection) spectrum of dichroic mirrors is sensitive to the angle of incidence of impinging light. After testing the angle dependence of several dichroic mirrors, we deduced that the cutoff wavelength is inversely proportional to the angle of incidence and surmised a rule of thumb whereby a change of ~1nm in the cutoff wavelength is achieved by ~$1^0$ change in the angle of incidence. Thus, increasing the angle of incidence by $1^0$ will result in a ~1nm blue shift in the cutoff wavelength. When we attempted to blue shift the cutoff for more than ~$5^0$, the transmission spectrum changed dramatically, while achieving a red shift proved easier and did not change the transmission spectrum significantly even when changing the angle of incidence by ~-$15^0$.

**Details of the nanorods synthesis**

ZnSe/CdS core-shell NCs was synthesized following a previously published procedure with some modifications.[1]

**Chemicals**

Hexadecylamine (HAD, 98%, Aldrich), diethylzinc (Et$_2$Zn, 1 M solution in hexane, Aldrich) Cadmium oxide (99.99%, Aldrich), Sulfur (99.999%, Aldrich), Selenium (99.999%, Aldrich), oleic acid (OA, 90%, Aldrich), trioctylphosphine (TOP, 90%, Aldrich), dodecylamine (98%, Fluka), trioctylphosphine oxide (TOPO, technical grade, 99% Aldrich), 1-octadecene (ODE, technical grade, 90% Aldrich), hexadecylamine (HPA, 99%, Aldrich), n-octadecylphosphonic acid (ODPA, 99%, PCI), methanol (anhydrous, 99.8%, Aldrich), hexane (anhydrous, 99.9%, Aldrich), toluene (99.8%, Aldrich). All chemicals were used as received without any further purification.

**Synthesis of ZnSe NCs**: 9.4 g of hexadecylamine was degassed under vacuum at 120 °C in a reaction flask, under argon flow, the mixture was heated up to 310 °C. Then, a mixture of 1 mL

1.0 M selenium dissolved in trioctylphosphine, 0.8 mL diethylzinc and 4 mL trioctylphosphine, was quickly injected. The reaction was continued at a constant temperature of 270 °C for 25 min and then cooled to room temperature.

**Preparation of Cadmium and Sulfur Stock Solutions.** 0.034 M cadmium oleate was prepared by mixing 0.03 g (0.24 mmol) CdO in 0.6 mL oleic acid and 6.4 mL ODE. The solution was heated to 280 °C under argon flow with rigorous stirring until all of the CdO dissolved. 0.29 M S solution was prepared by adding 23.3 mg sulfur in 2.5 mL of dodecylamine at ∼40 °C.

**CdS Shell Synthesis.** For typical CdS shell coating, 1.1 g unprocessed ZnSe cores, 5.3 mL octadecene (ODE) were loaded into a 50 mL three-neck flask. The solution was degassed at 100 °C. After that the solution was heated to 240 °C under argon, a mixture of 0.6 mL of 0.034 mmol/mL cadmium oleate stock solution and 0.06 mL of 0.29 mmol/mL sulfur stock solution was injected continuously at 0.72 mL/h. After the injection was finished, the mixture was further annealed for 5 min at 240 °C and then cooled down to room temperature.

**ZnSe/CdS-CdS NRs Synthesis.** This synthesis was adapted from the previously reported procedure in the literature.[2,3] In a typical synthesis CdO (60 mg), ODPA (290 mg) and HPA (80 mg) are mixed in TOPO (3.0 g). The mixture is degassed under vacuum at 150 °C for 90 min. After degassing step, the solution was heated to 380 °C under argon until it became clear, then 1.8 mL of TOP was injected and the temperature was recovered to 380 °C. Subsequently a solution of 120 mg S in 1.8 mL TOP mix with 40 nmol ZnSe/CdS nanocrystals is rapidly injected. Concentration of the ZnSe/CdS solution was estimated by the extinction at 436nm through a 1cm cuvette.[4] Then the growth was stopped after nanorods grow for 8 min at 365 °C. The NRs were precipitated with methanol and dispersed in toluene.

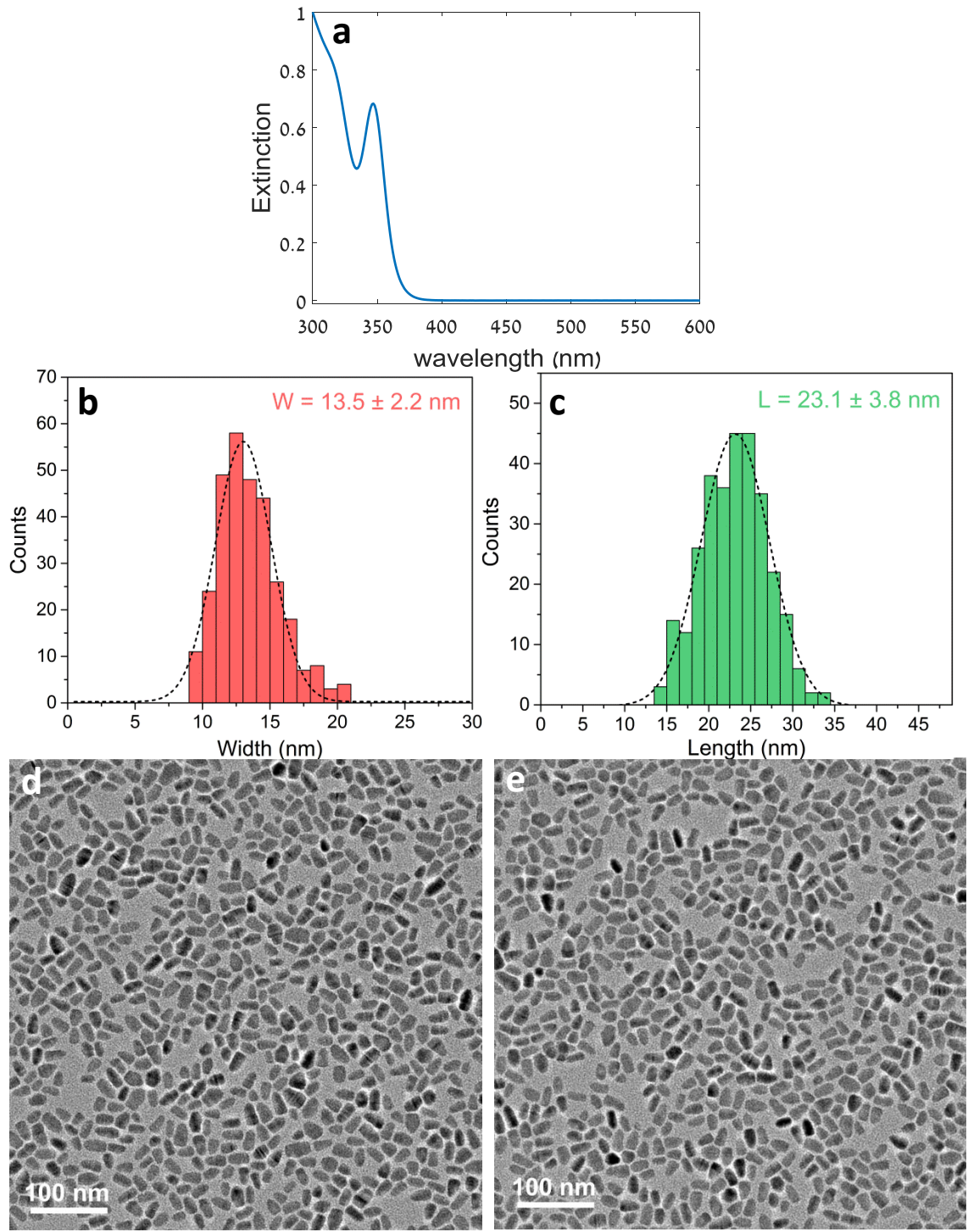

Figure S1 – (a) Extinction spectrum of ZnSe seeds. (b) width size distribution of the ZnSe/CdS NRs. (c) length size distribution of the ZnSe/CdS NRs. (d) and (e) TEM images of the final NRs.

NR's size was determined using ImageJ sampling over 200 different NRs and fitting a Gaussian to the histograms.

**Details of the analysis**

The blinking histogram is used to determine a threshold in order to eliminate dark "off" state periods from the QCSE response analysis.

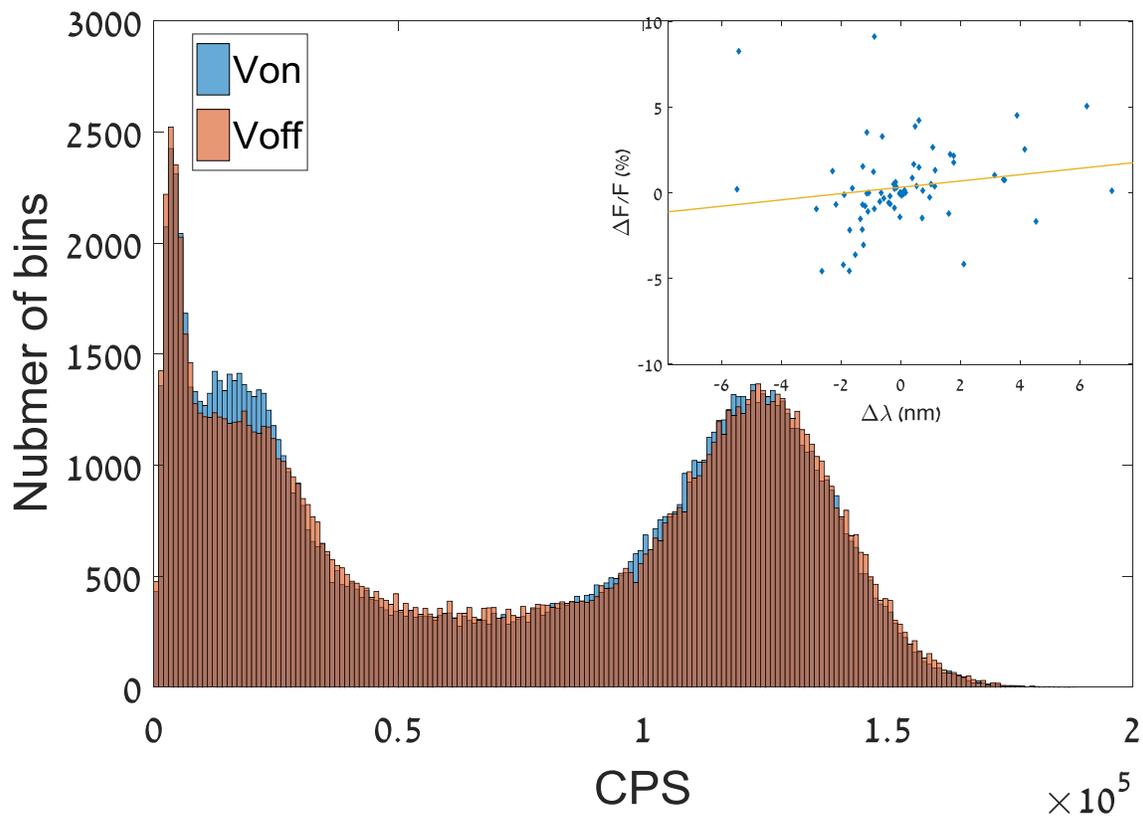

**Figure S2 – Example of a histogram over the blinking trace separated into $V_{on}$ an $V_{off}$ periods, it is clear this particle exhibits a small (-0.7%) intensity decrease as a response to the applied voltage. Inset shows a correlation between the intensity change and spectral shift.**

An example of the QCSE spectral shift dependence on the applied electric field:

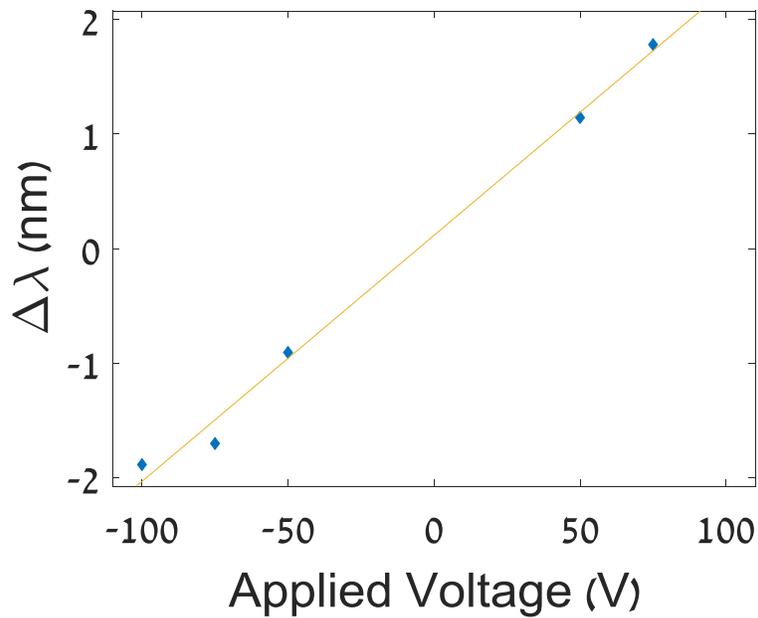

**Figure S3** – A single particle exhibiting a linear dependence of the spectral shift on the external electric field applied to it. A linear trend line is provided as a guide.

Determining whether a spectral shift occurred due to QCSE does not require specific knowledge of the dichroic cutoff wavelength, nor does it require a simulation for comparison. These are provided here as methods for unit conversion from the chosen unit-less estimator to nanometers. The conversion enables comparison of our results with previous work in the field, as the common and intuitive way to quantify the spectral shift is in nanometers or meV.

Estimating the spectral shift in nanometers by comparing real measurements to simulation requires two main assumptions. First, the transmission and reflection spectrum of the dichroic mirror used to split the emission peak is approximated by a step function with zero loss. The mirror's cutoff wavelength is taken from the measured spectrum of the dichroic. The transmission spectrum of the dichroic measured at a specific angle of incidence is given in Figure S1a together with a step function at 604nm. One can appreciate this approximation is reasonable. Second, the emission spectrum is approximated by a Gaussian function, the width of which is taken from measured single particle spectra (Fig. 2d).

Since this model may be written analytically it is possible to numerically solve. An alternative approach is to generate the objects in simulation and calculate from them the desired quantities. The analytic approach is described here in detail.

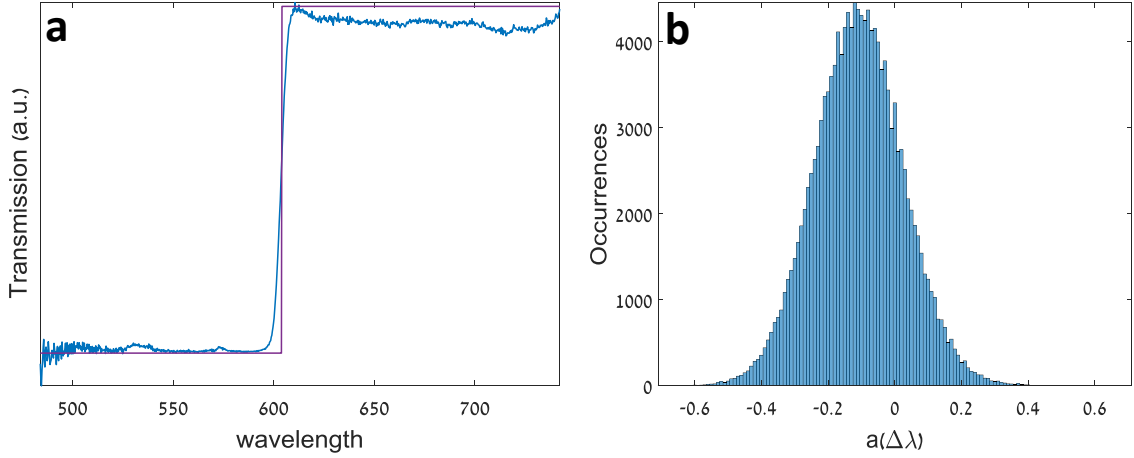

**Figure S4** – (a) Transmission spectrum of the dichroic used in the detection setup at a specific angle of incidence and a step function at 604nm. (b) A histogram of the shift estimator with binning of the voltage cycle period (1ms), the mean of the distribution is taken as the shift estimator reported in Figure 3c and the standard deviation is taken as the error for the Allan deviation analysis.

From the measured data, we extract the quantity $a_i = \dfrac{t_i}{t_i + r_i}$ separately from time bins when no voltage was applied (*0*) and those in which voltage was applied (*V*). Where, $t_0$ and $r_0$ are respectively the transmitted and reflected intensity when no voltage was applied and $t_V$ and $r_V$ are the same but for periods in which a voltage was applied. To relate this estimator to wavelength in nanometers we define:

$$g_0(x) = \dfrac{N}{\sigma\sqrt{2\pi}} e^{-\dfrac{(x-x_0)^2}{2\sigma^2}} , \quad g_V(x) = \dfrac{N}{\sigma\sqrt{2\pi}} e^{-\dfrac{(x-x_V)^2}{2\sigma^2}}$$

Where *x* is the wavelength, $g_0$ and $g_V$ are the Gaussian emission spectrum when no voltage was applied ($g_0$) and when voltage was applied ($g_V$), with central wavelengths $x_0$ and $x_V$ respectively, $N$ is the amplitude of the Gaussian corresponding to the intensity or, number of photons. σ, the standard deviation of the Gaussian function, is estimated from single particle spectra and set to 25nm. This is likely a minor overestimation of the width since single particle spectra require long exposures making broadening by spectral diffusion evident. We estimated the broadening of the Gaussian due to spectral diffusion as ~5nm on the time scale of the spectrum acquisition time. While this broadening may skew the results, its small contribution renders it negligible.

We further define $f(x)$ as the step function approximating a dichroic mirror with a cutoff wavelength $x_s$:

$$f(x) = \begin{cases} 0 & x < x_s \\ 1 & x > x_s \end{cases}$$

Thus we can calculate the intensity of transmission and reflection from the dichroic:

$$r_i = \int_{-\infty}^{\infty} (1-f(x)) g_i(x) dx = \int_{-\infty}^{x_s} g_i(x) dx$$

$$t_i = \int_{-\infty}^{\infty} f(x) g_i(x) dx = \int_{x_s}^{\infty} g_i(x) dx$$

Substituting and solving in the definition for $a_\Delta$:

$$a_\Delta = \frac{t_V}{r_V + t_V} - \frac{t_0}{r_0 + t_0} = \frac{1}{N}\left(\int_{x_s}^{\infty} g_V(x) dx - \int_{x_s}^{\infty} g_0(x) dx\right) =$$

$$= \frac{1}{2}\left(erfc\left(\frac{x_s - x_1}{\sqrt{2}\sigma}\right) - erfc\left(\frac{x_s - x_0}{\sqrt{2}\sigma}\right)\right)$$

To extract the spectral shift $(\Delta\lambda = x_1 - x_0)$ in nanometers one needs to estimate $a_0$ as well. Substituting in the definition for $a_i$ and solving for $x_0$ (for simplicity we assume $I_V = I_0$):

$$a_0 = \frac{t_0 - r_0}{I_0} = 1 - erfc\left(-\frac{x_s - x_0}{\sqrt{2}\sigma}\right)$$

$$\Rightarrow x_0 = x_s + \sqrt{2}\sigma erfc^{-1}(1 - a_0)$$

Extracting the difference:

$$\Delta x = \sqrt{2}\sigma \cdot \left(erfc^{-1}(2a_0) - erfc^{-1}(2(a_\Delta + a_0))\right)$$

The dichroic cutoff defines the sensitivity by influencing $a_V$ and $a_0$, as the shifts take the emission central wavelength away from the dichroic cutoff the sensitivity is diminished; this is illustrated in Figure 3b.

Using the experimental data, we extract a Δx value for each time bin of 0.5ms to get a series of values for $x_0$ and $x_V$. In the main text we present an estimator for the shift and not for the separate central wavelengths since it is not necessary to convert these results to nanometers for the purpose of detecting an applied voltage.

**Error estimation**

We present here the error estimation for the estimator presented in the main text for the spectral shift. Note, that the effects of spectral diffusion are neglected in this calculation. Assuming shot noise as the only noise source, that is: $\Delta t = \sqrt{t}$ and $\Delta r = \sqrt{r}$, and using the error propagation formula we describe the error in the estimator:

$$\Delta a_\Delta = \sqrt{\left(\Delta t_V \frac{\partial a}{\partial t_V}\right)^2 + \left(\Delta r_V \frac{\partial a}{\partial r_V}\right)^2 + \left(\Delta t_0 \frac{\partial a}{\partial t_0}\right)^2 + \left(\Delta r_0 \frac{\partial a}{\partial r_0}\right)^2}$$

Since intensity changes due to voltage are mostly negligible, and for simplicity, we assume: $\Delta F = 0 \Rightarrow I_V = I_0$:

$$\Delta a_\Delta = \sqrt{\frac{1}{I^3}\left(t_V r_V + t_0 r_0\right)} = \frac{1}{\sqrt{I}} \sqrt{\alpha_V(\lambda)(1-\alpha_V(\lambda)) + \alpha_0(\lambda)(1-\alpha_0(\lambda))}$$

Where $\alpha_i(\lambda) = \frac{r_i}{I}$ is a constant. From here we see that the error in $a(\Delta\lambda)$ scales as $N^{-0.5}$, where $N$ is the number of photons. This is used to calculate the error shown in Figure 3d by setting $N$ according to the count rate in the measurement.

It is important to note that bins are averaged over only if they are found consecutively in the "on state". For this reason, NRs exhibiting fast transitions with few prolonged "blink on" periods would have insufficient statistics for extended averaging.

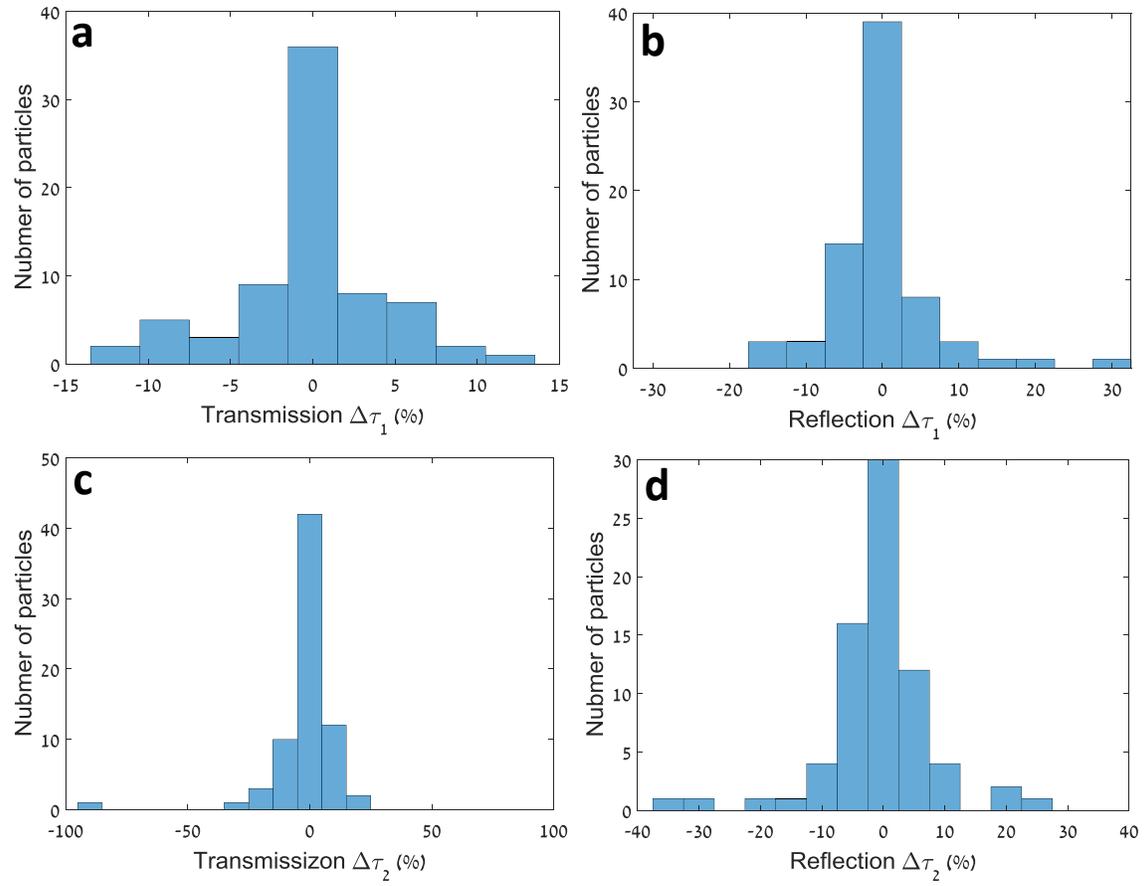

Figure S5 – Histograms of the relative lifetime variations. Δτ$_1$ and Δτ$_2$ refer, respectively, to the short and long lifetime components. Transmission and reflection refer to the detection channel.

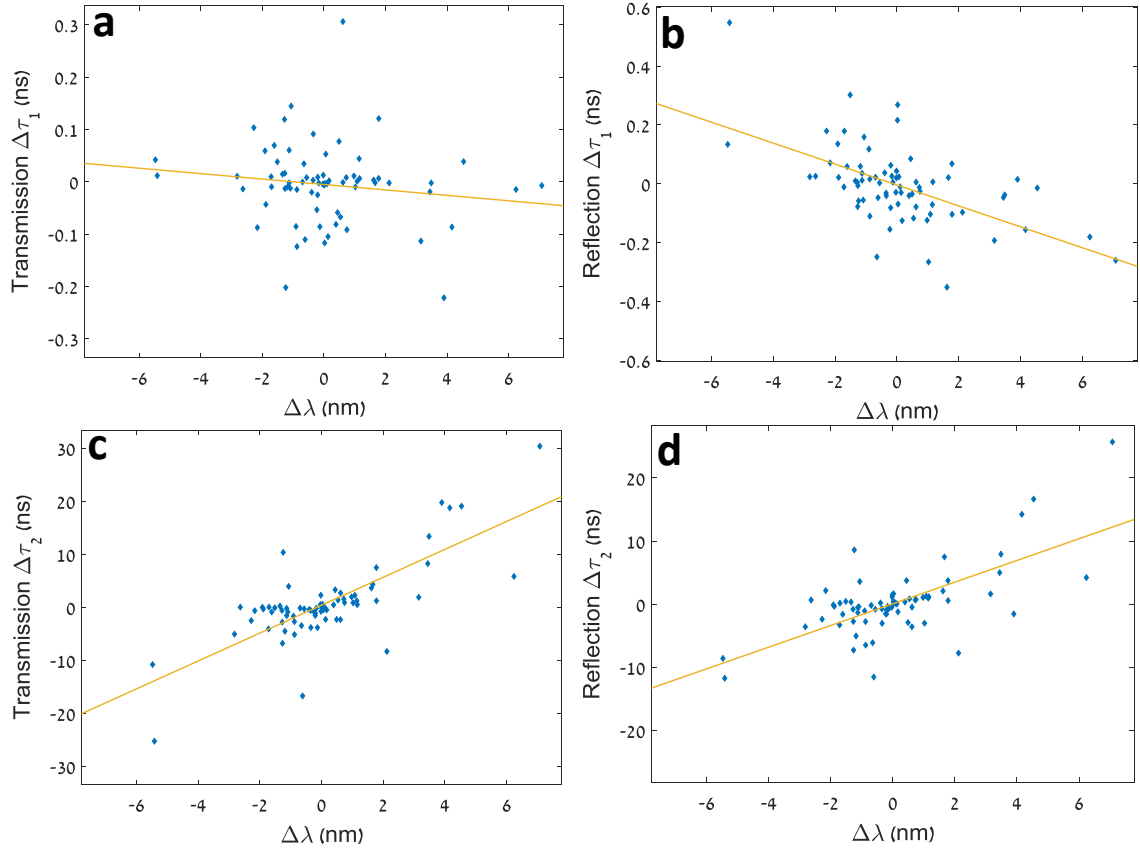

**Figure S6** – Scatter plots depicting the correlation between lifetime variations and spectral shifts due to QCSE. $\Delta\tau_1$ and $\Delta\tau_2$ refer, respectively, to the short and long lifetime components. Transmission and reflection refer to the detection channel.

We present the detection and false positive probabilities of a 1.5ms square voltage pulse calculated from the data collected from the NR which exhibited the largest QCSE spectral shift ($\Delta\lambda$). A threshold is set in units of σ, the standard deviation of $a(\Delta\lambda)$. One should consider that several NRs may be embedded in a single neuron to significantly increase the detection probability and nearly eliminate the false positive probability. For example, three well-oriented NRs would yield a false positive probability of 0.1% while maintaining a 61% detection probability, when considering a detection event when at least two of the three NRs crossed the detection threshold.

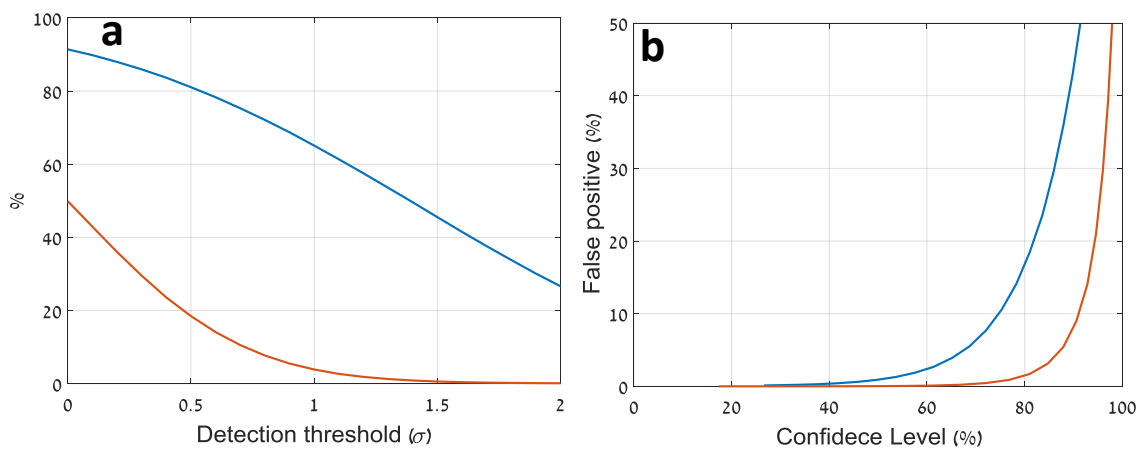

Figure S7 – Plots generated using the NR which exhibited the largest Δλ. (a) The detection confidence level (blue) and false positive probability (red) as a function of the threshold in units of σ. (b) The false positive probability as a function of the detection confidence level for when a single (blue) or three (red) NRs are considered.

**Table summarizing the measured NRs**

| NR# | Electrode gap* (μm) | Voltage (V) | Electric Field** (kV/cm) | Intensity*** (cps) | Δλ (nm) | Channel 1 Δτ (ns) | Channel 2 Δτ (ns) |
|---|---|---|---|---|---|---|---|
| 1 | 3.5 | 100 | 285 | 3.53E+04 | -2.3 | -2.5 | -2.4 |
| 2 | 3.5 | 100 | 285 | 6.67E+04 | -0.6 | -16.7 | -11.5 |
| 3 | 3.5 | 100 | 285 | 1.02E+05 | 3.9 | 19.8 | -1.5 |
| 4 | 3.5 | 100 | 285 | 1.37E+05 | 2.1 | -8.3 | -7.8 |
| 5 | 3.6 | 100 | 275 | 6.09E+04 | -5.4 | -25.2 | -11.7 |
| 6 | 3.8 | 100 | 267 | 4.92E+04 | 0.6 | -2.3 | -3.6 |
| 7 | 4.1 | 100 | 243 | 5.37E+04 | -1.2 | 10.4 | 8.6 |
| 8 | 4.1 | 100 | 243 | 4.04E+04 | 0.5 | -2.3 | -2.9 |
| 9 | 4.1 | 100 | 243 | 4.07E+04 | 0.6 | 2.7 | -0.5 |
| 10 | 4.1 | 100 | 243 | 4.56E+04 | 4.5 | 19.1 | 16.6 |
| 11 | 4.1 | 100 | 243 | 6.52E+04 | 7.1 | 30.4 | 25.7 |
| 12 | 4.1 | 100 | 243 | 3.27E+04 | 3.4 | 8.2 | 5.0 |
| 13 | 4.2 | 100 | 236 | 3.49E+04 | -0.9 | -5.1 | -6.5 |
| 14 | 4.1 | 100 | 243 | 5.16E+04 | -1.3 | -0.6 | -0.8 |
| 15 | 4.2 | 100 | 236 | 8.57E+04 | -1.3 | -2.8 | -2.7 |
| 16 | 4.0 | 100 | 251 | 3.89E+04 | 0.1 | -0.5 | 0.3 |
| 16 | 4.0 | 100 | 251 | 4.74E+04 | 0.0 | -0.2 | 1.7 |
| 17 | 4.0 | 100 | 251 | 5.44E+04 | 0.0 | -0.1 | 0.4 |
| 18 | 4.0 | 100 | 251 | 3.67E+04 | 0.0 | 0.5 | -0.5 |
| 19 | 4.0 | 100 | 251 | 7.69E+04 | -0.3 | -3.8 | -3.0 |
| 20 | 4.1 | 100 | 243 | 1.62E+05 | -2.2 | -0.6 | 2.2 |
| 20 | 4.1 | 100 | 243 | 9.47E+04 | -1.1 | 3.9 | 3.6 |
| 21 | 4.1 | 100 | 243 | 6.88E+04 | 3.5 | 13.4 | 7.9 |
| 22 | 4.1 | 100 | 243 | 5.82E+04 | 4.2 | 18.8 | 14.2 |
| 23 | 4.1 | 100 | 243 | 6.73E+04 | -0.6 | -3.4 | -6.1 |
| 24 | 4.1 | 100 | 243 | 5.15E+04 | 1.0 | 2.0 | 1.1 |
| 25 | 4.1 | 100 | 243 | 4.58E+04 | 0.7 | 1.6 | 0.7 |
| 26 | 4.1 | 100 | 243 | 3.51E+04 | -1.1 | -1.6 | -1.3 |
| 27 | 4.1 | 100 | 243 | 4.82E+04 | -5.5 | -10.8 | -8.6 |
| 28 | 4.0 | 100 | 251 | 6.84E+04 | -0.6 | -0.3 | -0.8 |
| 29 | 3.8 | 50 | 133 | 1.46E+05 | -1.1 | -0.9 | -0.4 |
| 29 | 3.8 | -50 | -133 | 1.39E+05 | 1.2 | 0.5 | 1.2 |
| 29 | 3.8 | 75 | 200 | 1.37E+05 | -2.6 | 0.0 | 0.7 |
| 30 | 3.9 | 50 | 129 | 7.89E+04 | 0.0 | 0.6 | 0.2 |
| 31 | 3.9 | 50 | 129 | 8.74E+04 | -1.3 | 0.2 | 0.3 |
| 32 | 3.6 | 50 | 138 | 8.30E+04 | -1.3 | -6.8 | -7.3 |
| 33 | 3.6 | 50 | 138 | 3.73E+04 | -1.5 | -0.1 | 0.4 |
| 33 | 3.6 | 50 | 138 | 3.29E+04 | -1.9 | 0.0 | -0.1 |
| 34 | 3.8 | 50 | 133 | 1.50E+05 | 0.7 | 0.9 | 0.8 |
| 35 | 3.6 | 50 | 138 | 1.60E+05 | -0.4 | -0.4 | 0.2 |
| 36 | 3.6 | 50 | 138 | 1.30E+05 | 0.2 | -0.5 | -1.2 |
| 37 | 3.8 | 50 | 133 | 2.79E+05 | 1.1 | 2.3 | 1.3 |
| 38 | 3.8 | 50 | 133 | 1.21E+05 | -0.1 | -3.8 | -0.7 |
| 38 | 3.8 | -50 | -133 | 1.11E+05 | 0.0 | -0.8 | 1.5 |
| 39 | 3.6 | 50 | 138 | 1.34E+05 | 1.6 | 3.7 | 2.1 |
| 39 | 3.6 | -50 | -138 | 9.67E+04 | -1.6 | -0.9 | -1.6 |
| 40 | 3.6 | 50 | 138 | 1.67E+05 | -1.1 | -0.2 | -0.5 |
| 41 | 3.8 | 50 | 133 | 2.03E+05 | 0.1 | 0.0 | 0.1 |
| 42 | 3.8 | 50 | 133 | 1.32E+05 | -0.2 | -0.2 | -0.3 |
| 43 | 3.8 | 50 | 133 | 1.28E+05 | 1.8 | 7.5 | 3.7 |
| 44 | 3.8 | 50 | 133 | 1.53E+05 | -1.2 | -4.5 | -5.0 |
| 45 | 4.1 | 50 | 122 | 2.52E+05 | 0.4 | 0.5 | 0.4 |
| 45 | 4.1 | 75 | 182 | 2.27E+05 | 0.5 | 1.4 | 0.8 |
| 46 | 4.1 | 50 | 122 | 1.14E+05 | 1.0 | 0.9 | -3.0 |
| 47 | 4.2 | 50 | 118 | 1.37E+05 | 1.1 | 1.2 | 1.0 |
| 47 | 4.2 | -50 | -118 | 1.08E+05 | -0.9 | -1.7 | -1.1 |
| 47 | 4.2 | 75 | 177 | 1.34E+05 | 1.8 | 1.2 | 0.6 |
| 47 | 4.2 | -75 | -177 | 1.11E+05 | -1.7 | 0.0 | 0.2 |
| 47 | 4.2 | -100 | -236 | 5.94E+04 | -1.9 | -0.3 | -0.3 |
| 48 | 4.1 | -100 | -243 | 1.31E+05 | -2.8 | -5.0 | -3.6 |
| 48 | 4.1 | 100 | 243 | 1.33E+05 | 6.2 | 5.8 | 4.2 |
| 48 | 4.1 | 50 | 122 | 9.18E+04 | 3.2 | 1.9 | 1.6 |
| 49 | 4.0 | 50 | 125 | 1.43E+05 | -0.2 | -1.5 | -1.6 |
| 50 | 4.0 | 50 | 125 | 1.68E+05 | -0.9 | -2.7 | -2.8 |
| 51 | 4.0 | 50 | 125 | 1.60E+05 | 1.7 | 4.3 | 7.5 |
| 52 | 3.8 | 50 | 133 | 2.08E+05 | -1.7 | -4.1 | -3.3 |
| 53 | 3.9 | 50 | 129 | 1.62E+05 | -0.2 | -0.9 | -1.5 |
| 54 | 3.9 | 50 | 129 | 8.46E+04 | 0.0 | 2.3 | 1.1 |
| 55 | 3.9 | 50 | 129 | 1.90E+05 | 0.4 | 3.3 | 3.7 |
| 56 | 3.9 | 50 | 129 | 1.54E+05 | -0.4 | -0.6 | -0.7 |
| 57 | 3.9 | 50 | 129 | 1.40E+05 | -0.7 | -0.1 | -0.2 |
| 58 | 3.9 | 50 | 129 | 8.01E+04 | -0.2 | -1.2 | -0.9 |
| 59 | 3.9 | 50 | 129 | 6.99E+04 | 0.1 | -2.3 | 0.1 |

*Measurement error of the electrode gaps is ±0.15µm

**Estimated by approximation to an ideal plate capacitor.

***The average intensity during periods that pass the "off state" intensity threshold.

**Details of the microelectrode fabrication process**

The electrodes were designed using Layout Editor. A mask was prepared on using standard chromium on soda lime-glass substrate (Nanofilm) exposed in a laser-writer (Heidelberg instruments, µPG 101).

Cover slips were immersed and washed with acetone, isopropyl alcohol (IPA), and distilled water, dried under a nitrogen stream and left for 5min on a hot plate at $150^0$C. Photoresist (Shipley, S1813) was spin coated at 2000RPM with a 1000 (RPM/s) ramp for 40sec followed by a soft bake of 60sec at $115^0$C. Exposure was done using the prepared mask and a mask aligner (Karl Suss, MA 6/BA 6) Development (Shipley, MF319) was performed for 55sec followed by washing with copious amounts of distilled water. The cover slips were further treated using $O_2$ plasma ashing (Diener Electronics Pico-DHP.) A thin adhesion layer of ~3nm of Cr was evaporated onto the substrates using an evaporator (Selene ODEM), subsequently Au was evaporated to yield any desired thickness in the range of 150-300nm. A chilling mechanism was activated during the Cr deposition. Liftoff was done in acetone for 1 hour and mild brief sonication. Resulting in inter-electrode gap of 2.5-4.5µm.